\documentclass[english]{article}
\usepackage[T1]{fontenc}
\usepackage[latin9]{inputenc}
\usepackage{geometry}
\usepackage{array}
\usepackage{float}
\usepackage{graphicx}

\makeatletter

\providecommand{\tabularnewline}{\\}

\@ifundefined{definecolor}
 {\usepackage{color}}{}
\usepackage{epsfig}
\usepackage{epstopdf}
\makeatother

\makeatother

\usepackage{babel}

\begin{document}

\title{A Simplified Hierarchical Dynamic Quantum Secret Sharing Protocol
with Added Features}

\author{Sandeep Mishra$^{a}$, Chitra Shukla$^{b}$, Anirban Pathak$^{b,}$%
\thanks{anirban.pathak@jiit.ac.in%
}, R. Srikanth$^{c,d}$, Anu Venugopalan$^{a}$ }

\maketitle
\begin{center}
$^{a}$University School of Basic and Applied Sciences, GGS Indraprastha
University, Sector 16C, Dwarka, New Delhi-110075, India 
\par\end{center}

\begin{center}
$^{b}$Jaypee Institute of Information Technology, A-10, Sector-62,
Noida, UP-201307, India 
\par\end{center}

\begin{center}
$^{c}$Poornaprajna Institute of Scientific Research, Sadashivnagar,
Bengaluru- 560080, India 
\par\end{center}

\begin{center}
$^{d}$Raman Research Institute, Sadashivnagar, Bengaluru- 560060,
India. 
\par\end{center}
\begin{abstract}
Generalizing the notion of dynamic quantum secret sharing (DQSS),
a simplified protocol for hierarchical dynamic quantum secret sharing
(HDQSS) is proposed and it is shown that the protocol can be implemented
using any existing protocol of quantum key distribution, quantum key
agreement or secure direct quantum communication. The security of
this proposed protocol against eavesdropping and collusion attacks
is discussed with specific attention towards the issues related to
the composability of the subprotocols that constitute the proposed
protocol. The security and qubit efficiency of the proposed protocol
is also compared with that of other existing protocols of DQSS. Further,
it is shown that it is possible to design a semi-quantum protocol
of HDQSS and in principle, the protocols of HDQSS can be implemented
using any quantum state. It is also noted that the completely orthogonal-state-based
realization of HDQSS protocol is possible and that HDQSS can be experimentally
realized using a large number of alternative approaches. 
\end{abstract}

\section{Introduction \label{sec:Introduction}}

In 1984, Bennett and Brassard \cite{bb84} introduced the first protocol
for quantum key distribution (QKD). Since then a large number of alternative
protocols of unconditionally secure QKD have been proposed \cite{ekert,b92,vaidman-goldenberg,N09}
and various aspects of secure quantum communication beyond QKD have
been explored \cite{Hillery,ping-pong,lm05,dsqcqithout-max-entanglement,dsqcwithteleporta,QSDC new-cluster,QCS,HQIS}.
For example, a large number of protocols have been proposed for quantum
secure direct communication (QSDC) \cite{ping-pong,lm05,Long   and   Liu,for PP,CL},
deterministic secure quantum communication (DSQC) \cite{dsqcqithout-max-entanglement,dsqcwithteleporta,dsqc_summation,entanglement      swapping,Hwang-Hwang-Tsai,reordering1,the:cao and song,the:high-capacity-wstate},
quantum dialogue (QD) \cite{Ba An,qd}, etc. All these protocols differ
from each other in some specific features. While both DSQC and QSDC
are used for secure direct quantum communication, DSQC protocols require
the exchange of additional classical information for decoding of the
message, whereas no such classical information is required in QSDC.
It is not the purpose of the present work to elaborately discuss these
aspects of secure quantum communication. In this work,  we focus on
two new aspects of secure quantum communication that have been introduced
recently: (i) dynamic quantum secret sharing (DQSS) \cite{DQSS,DQSS-Liao-2,DQSS-Jia}
and (ii) hierarchical quantum secret sharing (HQSS) (\cite{HQIS}
and references therein). These two recently introduced aspects of
quantum communication are extremely relevant for practical applications
and requirements in real life communication scenarios. In dynamic
secret sharing there exists a boss usually referred to as Alice and
she has several agents (say Bob and Charlie). Alice shares a secret
with Bob and Charlie that they can recover with the help of each other.
So far this is equivalent to quantum information splitting or the
traditional quantum secret sharing protocol introduced by Hillery
et al. in 1999 \cite{Hillery}. What makes it dynamic in the protocols
proposed in Refs. \cite{DQSS,DQSS-Liao-2,DQSS-Jia} is the inclusion
of the feature that an agent can always be added or dropped in the
scheme. This has direct practical relevance in real life situations.
Consider a company where Alice is a sales manager and her agents (Bob,
Charlie, David, etc.) are salesmen. Now, depending on the sales, she
may like to add or drop agents. Also, an agent may choose to quit
for various reasons (e.g., illness, more lucrative offers from another
company). This freedom of being able to recruit new agents and letting
old agents quit is an essential requirement for all practical setups.
However, this  feature was missing in the traditional protocols of
quantum secret sharing. A protocol with this feature is referred to
as DQSS protocol. In 2013, Hsu et al. proposed the first protocol
of DQSS \cite{DQSS}. It drew a lot of attention from the quantum
cryptography community because of its practical relevance and almost
immediately after its publication Hsu et al.'s proposal of DQSS, was
criticized \cite{DQSS-Wang,DQSS-Liao-comment} and two new protocols
of DQSS \cite{DQSS-Liao-2,DQSS-Jia} were proposed. Specifically,
Wang and Li \cite{DQSS-Wang} performed a cryptanalysis of the DQSS
protocol of Hsu et al. \cite{DQSS} and showed that if the first and
the last agent colluded with each other, they can obtain the secret
key of the boss without including the other agents. Further, Liao
et al. \cite{DQSS-Liao-comment} have shown that the DQSS scheme of
Hsu et al. is not completely secure if the new agents adopted in the
scheme are not honest. Liao et al. also proposed a new protocol \cite{DQSS-Liao-2}
of DQSS which is free from the collusion attack \cite{DQSS-Wang}
and dishonest user's attack \cite{DQSS-Liao-comment}. Jia et al.
have also proposed a protocol of DQSS \cite{DQSS-Jia}. All these
protocols of DQSS \cite{DQSS,DQSS-Liao-2,DQSS-Jia} use different
types of quantum states. For example, Bell states and entanglement
swapping was used in \cite{DQSS}, star-like cluster states were used
in the protocol of Jia et al \cite{DQSS-Jia} and GHZ states were
used in  the protocol of Liao et al. \cite{DQSS-Liao-2}. In the following,
we show that DQSS can be implemented using any quantum state by identifying
that all protocols of QKD or DSQC can lead to DQSS. We arrive at this
conclusion by noting from our earlier result \cite{With chitra IJQI}
that every quantum state can be used to implement efficient protocols
of DSQC and QKD. Thus, in brief, we can conclude that the aspect of
dynamism in terms of addition and deletion of agents has been successfully
included in the recent papers on DQSS \cite{DQSS,DQSS-Liao-2,DQSS-Jia}
and it is possible to implement DQSS schemes using any arbitrary quantum
state. However, no organizational hierarchy among the agents was present
in the Hsu et al. protocol and it required Bell states. Another practically
relevant scheme of modified quantum secret sharing is HQSS introduced
by some of the present authors \cite{HQIS}. In HQSS all the agents
are not equally powerful thus an organizational hierarchy exists.
Several examples of important practical situations where use of HQSS
is essential are provided in our earlier paper \cite{HQIS}. However,
no investigation on the possibility of inclusion of new agents or
dropping of agents were performed in the previous study on HQSS. In
practical situations a realistic scheme of secret sharing should have
both the features as all existing organizations have an internal hierarchy
among the staff members and staff members have the right to take leave
or resign. In what follows, we aim to combine these two features (i.e.,
dynamism and hierarchy) and propose a new protocol of hierarchical
dynamic quantum secret sharing which will be immensely relevant for
all practical applications. Further, we will show that it is quite
easy to implement a protocol of dynamic quantum secret sharing as
every protocol of QKD, quantum key agreement (QKA), DSQC and QSDC
can be transformed to a protocol of hierarchical dynamic quantum secret
sharing.

The remaining part of the paper is organized as follows. In Section
\ref{sec:Simplified-protocol}, we propose a very simple protocol
of dynamic quantum secret sharing that has all the advantages of Hsu
et al.'s protocol of dynamic quantum secret sharing, but can be implemented
using any protocol of QKD, QKA, QSDC or DSQC as the backbone of the
proposed protocol. It is also shown that the proposed protocol has
an intrinsic hierarchy among the agents and thus it can be viewed
as a protocol of hierarchical dynamic quantum secret sharing. In Section
\ref{sec:Security-and other-features}, we discuss the security and
other features of the protocol and specifically show that the proposed
protocol can be modified to yield a protocol of controlled hierarchical
dynamic quantum secret sharing and it can also be used to communicate
meaningful information (instructions) among the agents. In Section
\ref{sec:Efficiency}, the proposed protocol is compared with the
existing protocols in terms of qubit efficiency (for convenience of
comparison, we ignore the hierarchical aspect here) and allowed features.
Our analysis shows that the proposed protocol can be implemented with
maximal efficiency. Finally, the paper is concluded in Section \ref{sec:Conclusions}.

\begin{figure}[H]

\begin{centering}
\includegraphics[scale=0.5]{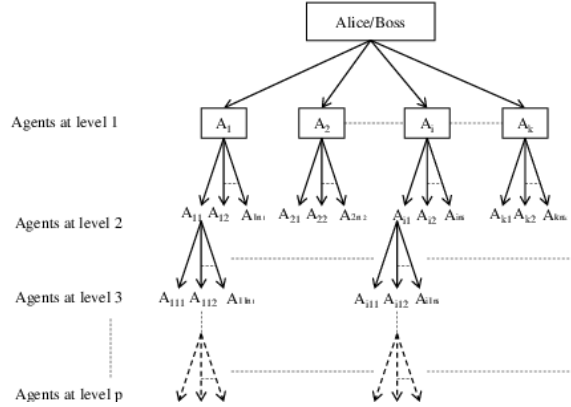} 
\par\end{centering}

\caption{\label{fig:tree} The hierarchical structure of the organization can
be viewed as a tree. The lowest level agents are the leafs of the
tree. Each agent is connected with his/her boss via a sub-protocol
which may be any protocol of QKD, DSQC, QSDC or QKA, but all sub-protocol
may not be composable. At least in case of BB84 protocol, proof of
both composability of this tree and unconditional security is available.
Indicating that BB84 protocol is a good choice of sub-protocol in
this tree structure.}
\end{figure}

\section{Simplified protocol of dynamic quantum secret sharing \label{sec:Simplified-protocol}}

Consider that Alice is the boss and she has $l$ primary agents with
whom she wishes to share a secret directly in a way that incorporates
the dynamism (as discussed in the previous section). 
\begin{description}
\item [{Step~1}] Alice and her $i$-th agent faithfully follow a specific
protocol of QKD/QKA/DSQC/QSDC to generate an $n$-bit symmetric key
$K_{A_{i}}=K_{i}$ ($i\in\{1,2,\cdots,l\}$) for secret sharing. For
example, if Alice has two agents Bob and Charlie then after following
the protocol independently with Bob and Charlie, Alice produces $K_{A_{1}}=K_{B}$
and $K_{A_{2}}=K_{C}$, where the subscripts $A,\, B,\, C$ stands
for Alice, Bob and Charlie, respectively. Subsequently, Alice produces
a master key $K_{M}=K_{A_{1}}\oplus K_{A_{2}}\oplus\cdots\oplus K_{A_{l}}.$
In the present case $K_{M}=K_{B}\oplus K_{C}.$ Clearly $K_{M}$ can
be recovered by the agents iff all of them collaborate with each other. 
\item [{Step~2}] If a new agent David wants to join the protocol then
he has two options. Either he can join directly with Alice and in
that case he will be referred to as a primary agent or he could join
a previously appointed primary agent (say Bob) as a secondary agent.
Without loss of generality, we consider the case that he joins with
Alice as the primary agent. In this case, Alice will faithfully follow
a specific protocol of QKD/QKA/DSQC/QSDC to generate an $n$-bit symmetric
key such that $K_{A_{l+1}}=K_{D}$. 
\item [{Step~3}] Alice will update her master key as $K_{M}^{\prime}=K_{M}\oplus K_{D}$.
This would automatically include David as a primary agent since after
updating of the master key by Alice, Bob and Charlie together will
not be able to obtain the secret of Alice, unless David collaborates
with them. In a similar manner agent Bob can collaborate with David
and recruit him as a secondary agent. 
\item [{Step~4}] If an agent, say David, wants to quit the scheme then
Alice (his immediate boss) would update her master key as $K_{M}^{\prime\prime}=K_{M}^{\prime}\oplus K_{D}$.
This operation will ensure that the secret of Alice can be recovered
by users (agents) other than David, if all of them collaborate. The
above description shows how a dynamic scheme of quantum secret sharing
can be implemented in an unconditionally secure manner using any protocol
that can be used for symmetric key distribution between two authenticated
users. It is interesting to note that here the agents of Alice do
not need to be quantum. All of them can be classical (i.e., restricted
to classical operations: measurement and preparation of quantum states
in the computational basis). This makes the protocol relevant in the
context of several semi-quantum protocols of QKD which have been recently
proposed (\cite{semi-quantum} and references therein), where the
schemes are implemented keeping Bob classical. The above protocol
clearly shows that a successful implementation of QKD is sufficient
for designing of a protocol of DQSS. Thus, by using a semi-quantum
QKD scheme in the \textbf{Step 1} of the above protocol we can easily
obtain a protocol for semi-quantum DQSS. Following the same logic,
we can state that as completely orthogonal-state-based implementation
of QKD and DSQC/QSDC are possible (\cite{ortho-review} and references
therein), it is possible to design completely orthogonal-state-based
protocol of HQDSS.
\end{description}

\subsection{Intrinsic hierarchy in the protocol \label{sub:Hierachy-in-the-protocol}}

The scheme presented above has an inherent hierarchy present inside
it which can be understood easily if we consider Alice as the boss
(master) whose key is $K_{M}$ and Bob, Charlie and David as agents
of Alice with the keys $K_{B},\, K_{C}\,{\rm and}\, K_{D}$, respectively.
Now let us suppose that a new agent, Elsa, joins the scheme such that
she faithfully follows a secure protocol with Bob, who is an agent
of Alice to share a symmetric key $K_{E}$ with him. After following
the protocol Bob can update his key as $K_{B}^{\prime}=K_{B}\oplus K_{E}$.
Now if Bob uses the key $K_{B}^{\prime}$ then all the agents Bob,
Charlie, David and Elsa would be required to cooperate with each other
to obtain the information shared by Alice. However, if Bob decides
to use the key $K_{B}$ then the agents Bob, Charlie and David can
bypass the agent Elsa to decode the secret of boss Alice. Thus a secondary
agent is less powerful than a primary agent and this illustrates the
intrinsic hierarchy present in this protocol of dynamic secret sharing.
Specifically, if an agent joins the scheme by following the protocol
with any agent other than Alice, then that agent's access to the information
shared by Alice will be at the mercy of his immediate boss (say, Bob
in the case of Elsa). Thus the new agent Elsa will be under the control
of her boss, Bob. This implies that if the above mentioned scheme
is used, then we can implement dynamism and hierarchy among the agents
with the help of any protocol capable of sharing an n-bit key between
two users.

\section{Security of the protocol and its additional features \label{sec:Security-and other-features}}

In the proposed protocol, it is obvious that the security of the protocol
is equivalent to the security of the scheme used to obtain keys between
a boss and his/her agents. For example, if Alice and Bob use BB84
(B92) protocol to obtain $K_{A}=K_{B}$ then the security proof of
BB84 (B92) would ensure the security of the present scheme. As BB84,
B92 and other single-particle based protocols can also be used to
implement dynamic quantum secret sharing, we may avoid the use of
Bell states and Bell measurement in Hsu et al. protocol as this would
reduce the requirement for quantum resources. Further, as we have
shown that all the existing protocols of QKA, QKD. DSQC and QSDC can
be turned into protocols equivalent to Hsu et al.'s protocol of dynamic
secret sharing \cite{With chitra IJQI}. and thus the present version
of the protocol provides many alternative ways of realization of dynamic
secret sharing.

\subsection{The collusion attack of the agents and the honesty check of a revoked
agent}

In the recent work of Liao et al. \cite{DQSS-Liao-2}, they have discussed
two possible security issues related to the DQSS protocol proposed
by them. Specifically, they discussed the security of their protocol
against the collusion attack of a subset of the set of all agents.
As in all the existing DQSS protocols the master key of the boss $K_{M}=K_{A_{1}}\oplus K_{A_{2}}\oplus\cdots\oplus K_{A_{l}}$
is obtained by a modulo 2 summation over the keys of all the agents
at level 1 (note that, except the present protocol of HDQSS, in all
other protocols of DQSS all the agents are in the level 1 only), any
collusion of $n$ agents with $n<l$ will never reveal $K_{M}$. Further,
in the proposed HDQSS, an agent of level $p$ may decide to include
some or all of his sub-agents in level $p+1$. Thus, in the present
protocol, it is allowed that all the agents of level 1 and all sub-agent
of agent $i$ who are located at level 2 cooperate to obtain the secret
of the boss, but any collusion attack of a proper subset of the set
of these agents will never lead to $K_{M}$ and consequently the present
protocol in particular and all the existing DQSS protocols in general
are free from collusion attack of a subset of agents. Further, Liao
et al. discussed a complex revocation process, which in our opinion
is not technically necessary. Technically, in our protocol a copy
of the key of a particular agent belonging to level $k$ is available
with his immediate boss who is either an agent in level $k-1$ or
the ultimate boss. Thus, all that needs to be done now is that the
immediate boss of the agent to be removed refreshes his key by performing
an XOR operation of his key with the key of the agent to be removed.

\subsection{Composability and related issues \label{sub:Composability}}

The key insight in this work is the fact that the security of hierarchical
quantum secret sharing schemes can be reduced to security of bipartite
QKD. Note that similar ideas of reduction in the cryptographic context
for orthogonal state based encoding is already discussed by some of
the present authors (\cite{1-Chitra-under-prep} and references therein).
The main requirement for the present reduction to work is that the
bipartite QKD used must be composable \cite{2-Wehner}. The issue
of composability arises when we wish to build a complicated, composite
cryptographic protocol that uses simpler cryptographic primitives
as subroutines. Thus, composability issue arises in the context of
the proposed HDQSS protocol as from Fig. \ref{fig:tree}, it is clearly
evident that our protocol of HDQSS consists of a tree of sub-protocols.
Specifically, the cryptographic protocol followed by Alice and her
$i$-th agent to share a key $K_{A_{i}}$ in the Step1 of our protocol
is a sub-protocol. Similarly, the protocol used by the $j$th agent
of level $p-1$ and his/her $k$-th agent (who is in level $p$) is
a sub-protocol. Such sub-protocols that are present in a tree are
referred to as the children of the tree. The primitives are called
leaves. In our case all the agents of the lowest level are leaves
of the tree. Similar argument related to the arise of composability
issue is applicable to all the existing DQSS protocols. However, composability
of the DQSS protocols are not discussed in any of the existing papers.
The issue is important because a particular subprotocol may have stand-alone
security, but may leak some information that could be harmful when
it is part of a larger, parent protocol. Thus we require composability
over and above the stand-alone security of the subprotocols. 

The security of a complex cryptographic protocol (tree) is established
by first establishing the security of the primitives and then using
that result to obtain the security for the parent sub-protocols, and
so on until one reaches the root of the tree (Alice/boss in our case)
\cite{composability1,compsability2}. This bottom-up approach is used
to establish the security of the earlier proposed DQSS protocol and
the same approach is adopted above to establish the security of our
protocol. However, not all sub-protocols are expected to provide composability.
A good choice of sub-protocol for the sharing of a key from an agent
to his sub-agents may be the BB84 protocol%
\footnote{All bi-partite QKD protocols are composable \cite{composability1},
but we prefer BB84 over other protocols because in case of BB84 clear
proof unconditional security \cite{Shor-Preskill} and strict upper
limit of the tolerable noise is known.%
} as it is composable \cite{composability1} and unconditionally security
\cite{Shor-Preskill}. Here it would be apt to note that the idea
of sequential composability was introduced in Ref. \cite{3-canetti}
in the context of secure multi-party computation. In sequential composition,
one allows primitive protocols to be composed arbitrarily provided
that at any given time, precisely one instance of the protocol is
being executed. Any other instance of the protocol can be executed
only when the present instance halts. A stronger form of security
is required for universal composability, in which the protocols being
composed are allowed to run concurrently \cite{4-backes,5-Canetti}. 

In composable security for the simulation paradigm, we require that
the environment is unable to distinguish between the real protocol
primitive from its ideal black box functionality. If this were not
the case, then an adversary could potentially use environmental information
about a previous run of the protocol embedded in the parent protocol.
This information that the adversary receives from the environment
could be in the form of a quantum state from the environment or entanglement
with the environment. The general criteria for composable security
in the quantum and classical contexts are presented in \cite{compsability2,7-unruh}.
In the present work, we require universal composability of unconditionally
secure quantum key distribution, which is indeed known \cite{composability1}.
That is, it is shown in \cite{composability1} that QKD (i.e., usual
security definition for QKD) also entails composable security. This
means that the key generated by a QKD subroutine in the hierarchy
can indeed be used subsequently. As a consequence the protocol of
HDQSS presented here is composable in all such situations where the
subprotocols used are the protocols of QKD.

\subsection{Promotion of an agent\label{sub:Promotion}}

In a practical organizational scenario, we need a hierarchy among
staff and dynamism in the organization in the sense that a new staff
can join or an old staff can resign or be terminated. The existence
of these desirable features in our protocol is already discussed.
However, in a realistic situation we need another feature: the possibility
of growth of an employee. To be precise, an agent who is performing
well in level $l>1$ must have some option to be promoted to the level
$l-1$. The dynamic nature of the proposed HDQSS protocol automatically
ensures that as it allows the agent at level $l$ to resign from his
present job and as it also allows another agent in level $l-2$ to
recruit him as a new agent of level $l-1.$

\subsection{Sending a meaningful classical message using the protocol \label{sub:Sending-a-meaningful}}

In the existing protocols of dynamic quantum secret sharing only a
key is shared among agents of Alice as the master key $K_{M}$ is
generated via probabilistic outcomes of the measurement. The same
would be the case here if we use a protocol like BB84 or any protocol
of QKA in Step 1 of our protocol. However, it is straightforward to
understand that at a later time (after the key $K_{M}$ is shared
among the agents as $K_{M}=K_{B}\oplus K_{C}\oplus K_{D}+\cdots)$
Alice can use the key to send an instruction (information) to her
agents using the key. As Kerckhoff's principle defines that a message
is secure when encrypted with a secure key, the unconditional security
of the shared key would implement unconditional security of the shared
message. To be more precise, if Alice wishes to send a message $M_{A}$
she may publicly announce $S_{A}=K_{M}\oplus M_{A}$ and subsequently
all her agents can collaborate to obtain $M_{A}=S_{A}\oplus K_{M}=S_{A}\oplus K_{B}\oplus K_{C}\oplus K_{D}+\cdots$.

\subsection{Turning the protocol into a protocol of controlled secret sharing\label{sub:Turning-the-protocol-into-a protocol of conntrolled}}

Now it may be of interest for Alice to share the key at some time,
but not to allow her agents to collaborate and produce the key till
she desires that the agents do so. For example, consider that the
President of a country (Alice) shares a key that would require opening
the switch to a nuclear weapon among chiefs of army, navy and air
force, but the president does not want that the agents collaborate
among themselves and open the weapon at a time not desired by her.
In such a situation, Alice would require to have a control over the
key. This control may be ensured in several ways. For example, Alice
may create a shared key with all her agents as follows $K_{A_{1}}=K_{B},K_{A_{2}}=K_{C},K_{A_{3}}=K_{D}$
and instead of creating a master key $K_{M}$ as $K_{M}=K_{B}\oplus K_{C}\oplus K_{D}+\cdots$
she may create the master key as $K_{M}=K_{B}\oplus K_{C}\oplus K_{D}^{\prime}+\cdots$
where $K_{D}^{\prime}=\Pi_{n}K_{D}$ and $K_{i}$ is a $n$-bit sequence
and $\Pi_{n}$ is a permutation operator that randomly permutes an
$n$-bit sequence. Now, as $\Pi_{n}$ is unknown to David, he does
not know $K_{D}^{\prime}$ and consequently Bob, Charlie and David
are not allowed to obtain $K_{M}$ till Alice allows them to do so
by disclosing the detail of permutation operations applied by her
(say, when she considers it is the right time to open the nuclear
weapon).

\section{Efficiency of the proposed protocol\label{sec:Efficiency}}

Efficiency of quantum communication protocols is calculated using
two analogous but different parameters. The first one is simply defined
as \begin{equation}
\eta_{1}=\frac{c}{q},\label{eq:efficency 1}\end{equation}
where $c$ is the total number of classical bits (message bits) transmitted/shared
using the protocol and $q$ denotes the total number of qubits used
for the purpose \cite{the:C.-W.-Tsai,hwang-hwang-Tsai}. This measure
has been used by Liao et al \cite{DQSS-Liao-2} in their recent work
on DQSS \cite{DQSS-Liao-2} to establish that their protocol is more
efficient than the earlier proposed DQSS protocol of Hsu et al. \cite{DQSS}
and Jia et al. \cite{DQSS-Jia}. Their claim is not completely correct.
Before we illustrate this point let us find out an upper bound on
$\eta_{1}$ for DQSS protocols. For convenience of comparison with
the earlier works we consider an $m$-party DQSS scheme with Alice
as the boss having $(m-1)$ agents. Now Alice has to implement a sub-protocol
with each of these agents. The maximum efficiency for each of these
sub-protocols can be $\frac{1}{2}$ \cite{anindita-pla}. This is
so because if $2x$ qubits (consider a random mix of verification
qubits and message qubits) travel through a quantum channel accessible
to Eve and the possibility of eavesdropping is checked by using $x$
of them, then for any $\delta>0,$ the probability of obtaining less
than $\delta n$ errors on the verification qubits, and more than
$(\delta+\epsilon)n$ errors on the remaining $x$ qubits is asymptotically
less than $\exp[-O(\epsilon^{2}x)]$ for large $x$ \cite{anindita-pla,nielsen}.
Thus to ensure the unconditional security of the sub-protocol operating
between Alice and her $i$th agent of level 1, it is required that
they (Alice and her specific agent) check half of travel qubits for
eavesdropping. Thus, $\eta_{1max}=\frac{1}{2}$ and it is easy to
recognize that in $m$-party DQSS, Alice prepares a 1-bit secret or
key $K_{M}=K_{A_{1}}\oplus K_{A_{2}}\oplus\cdots\oplus K_{A_{m-1}}$
by combining all the $1$ bit secrets that she shares with each of
the agents and consequently she needs $2(m-1)$ qubits to create a
single bit of secret ($K_{M}$). Equivalently, she requires $m-1$
sub-protocols of efficiency $\eta_{1}=\frac{1}{2}.$ In brief, upper
bound on $\eta_{1}$ of an unconditionally secure DQSS is $\frac{1}{2(m-1)}$.
This bound can be achieved by using different sub-protocols as in
our earlier works where we have described a large number of protocols
with $\eta_{1}=\frac{1}{2}$ (cf. \cite{With chitra IJQI,anindita-pla,chitra-w-state}).
In what follows we have assumed that in the DQSS protocol proposed
here one of the maximally efficient QKD or QSDC protocol proposed
in Refs. \cite{With chitra IJQI,anindita-pla,chitra-w-state} is used
as sub-protocols and consequently $\eta_{1}$ for our protocol is
$\frac{1}{2(m-1)}$.

The simple measure described above (\ref{eq:efficency 1}) does not
include the classical communication that is required for decoding
of information in case a DSQC protocol is used as the sub-protocol.
Further, for implementation of any DQSS scheme one of the users will
finally recover the secret of the boss and for that he/she would require
the help of other agents. Thus, in the implementation of an $m$-party
DQSS $m-2$ agents must send one bit of information to the agent responsible
for the recovery of the secret of Alice (boss). Consequently, even
if we apply a QKD/QSDC sub-protocol we need an additional $m-2$ bits
of classical information for final decoding of the 1 bit secret key
of the boss. As $\eta_{1}$ does not include these classical bits,
so it may be considered as a weak measure. There exists another measure
of efficiency \cite{defn of qubit efficiency} that is frequently
used and includes the classical communication and is given as \begin{equation}
\eta_{2}=\frac{c}{q+b},\,\label{eq:efficiency 2}\end{equation}
where $b$ is the number of classical bits exchanged for decoding
of the message (classical communications used for checking of eavesdropping
are not counted). So for our protocol $b=m-2$ and consequently the
maximum value of $\eta_{2}=\frac{1}{(2m-2)+(m-2)}=\frac{1}{3m-4}$.
Similarly, one can obtain values of $\eta_{2}$ for other existing
protocols of DQSS \cite{DQSS,DQSS-Liao-2,DQSS-Jia}, too. However,
we have not tried that here as a comparison of the efficiencies of
the existing protocols of DQSS has already been presented in the recent
work of Liao et al. \cite{DQSS-Liao-2} using $\eta_{1}$. Following
them we and compare our protocol with the existing protocols in Table
\ref{tab:Comparison}. From the third row of the Table \ref{tab:Comparison},
we can easily observe that for three-party DQSS, our protocol is more
efficient than Liao et al.'s protocol which in turn is more efficient
compared to Hsu et al.'s protocol. Liao et al. used this merit of
their protocol to establish their protocol as superior to the protocol
of Hsu et al. in terms of efficiency. However, the benefit of better
quit efficiency disappears for asymptotically large $m$ values. To
specifically elaborate this point in the next row of the Table \ref{tab:Comparison},
we have provided $\eta_{1}$ for all the protocols for a 50-party
DQSS. We can easily see that for 50-party DQSS efficiency of our protocol,
and that of Liao et al. protocol \cite{DQSS-Liao-2} and Hsu et al.
protocol \cite{DQSS} is practically the same. Apart from achieving
the maximum possible efficiency, the proposed protocol has some more
advantages over the existing protocols. These advantages are summarized
in the last two rows of Table \ref{tab:Comparison}.

\begin{table}[H]

\begin{centering}
\begin{tabular}{|>{\centering}p{1in}|>{\centering}p{1in}|>{\centering}p{1in}|>{\centering}p{1in}|>{\centering}p{1in}|}
\hline 
 & Hsu et al. protocol \cite{DQSS}  & Jia et al. protocol \cite{DQSS-Jia}  & Liao et al. protocol \cite{DQSS-Liao-2}  & Proposed protocol

(using a maximally efficient sub-protocol of QKD or QSDC described
in Refs. \cite{With chitra IJQI,anindita-pla,chitra-w-state} \tabularnewline
\hline 
Qubit efficiency $\eta_{1}$ ($m$-party DQSS)  & $\frac{1}{2m}$  & $\frac{1}{4m-2}$  & $\frac{1}{2m-1}$  & $\frac{1}{2m-2}$\tabularnewline
\hline 
Qubit efficiency $\eta_{1}$ ($3$-party DQSS)  & 16.67\%  & 10\%  & 20\%  & 25\%\tabularnewline
\hline 
Qubit efficiency ($50$-party DQSS)  & 1\%  & 0.51\%  & 1.01\%  & $1.02$\%\tabularnewline
\hline 
Requirement of quantum entanglement  & Essential as uses Bell state  & Essential as uses cluster state  & Essential as uses GHZ state  & Not essential as it can be implemented using single qubit state\tabularnewline
\hline 
Features  & Only dynamic  & Only dynamic  & Only dynamic  & Dynamic and hierarchical; can be implemented using any protocol of
QKD, QD, QKA, DSQC, QSDC, etc.; an agent can be promoted to the next
level.\tabularnewline
\hline
\end{tabular}
\par\end{centering}

\caption{\label{tab:Comparison}Comparison of the proposed protocol with the
existing protocols \cite{DQSS,DQSS-Liao-2,DQSS-Jia} }
\end{table}

\section{Conclusions\label{sec:Conclusions}}

To conclude, in this paper, we have proposed a simplified protocol
of HDQSS. The protocol is interesting for several reasons. Firstly,
it includes features from recently introduced ideas of HQSS and DQSS.
Secondly, it can be implemented by using a large number of alternative
protocols of secure quantum communication as sub-protocols. More specifically,
the proposed protocol can be implemented using any existing protocol
of QKD, QSDC, DSQC or QD. Further, it is also possible to design a
semi-quantum protocol of HDQSS and in principle the protocols of HDQSS
can be implemented by using any quantum state as in Ref. \cite{With chitra IJQI}
we have already established that any arbitrary quantum state can be
used to implement maximally efficient protocols of QKD, DSQC and QSDC.
Finally, the protocol has some features (e.g., hierarchy in the organization,
dynamism, possibility of promotion of the agents, etc.) that were
not simultaneously present in any of the existing protocols of related
tasks. Further, in this work we have also discussed security of the
proposed protocol against eavesdropping and collusion attack with
special attention to the issues related to composability which is
extremely relevant to the complex protocols of this kind that are
essentially built by using several sub-protocols. The efficiency of
the proposed protocol is compared with that of other existing protocols
of DQSS and it is shown that the presented protocol has better efficiency
when a small number of agents are involved, but the efficiency is
practically the same as that of the protocols of  Hsu et al. and Liao
et al. when a large number of agents are present. 

Here it would be apt to note that in the present paper we have assumed
that the immediate boss of an agent keeps a copy of the key that he/she
shares with that particular agent and thus the boss enjoys the privilege
of kicking out an agent at her whim. Interestingly, some security
reasons may lead to a situation where the immediate boss of the agent
is not allowed to store the key that he shares with the agent. This
would lead to a very different scenario. Specifically, in this situation
the boss would require the agent to give his/her key in order to eliminate
him/her, thus bringing in his integrity into the picture. However,
if we assume that the station of Alice and each of the agents (i.e.,
all the nodes in the graph shown in Fig. \ref{fig:tree}) are secure,
then classical information (shared key) can be securely stored and
the above said integrity of the agent will not be required. Further,
it may be noted that the completely orthogonal-state-based realization
of HDQSS protocol is also possible as completely orthogonal-state-based
realization of QKD, QSDC and DSQC are possible \cite{vaidman-goldenberg,With chitra IJQI,ortho-review}.
Finally, as the proposed protocol is extremely relevant for various
practical situations and it is possible to implement it using various
alternative sub-protocols that are already experimentally realized,
we hope that experimentalists will find it interesting to implement
this HDQSS scheme and the idea discussed here will substantially contribute
to the development of future implementations of secure quantum communication
schemes.

\textbf{Acknowledgment:} A. P. thanks the Department of Science and
Technology (DST), India, for the support provided through DST project
No. SR/S2/LOP-0012/2010.

\end{document}